# NANOCOMPOSITE Si-C-N COATINGS

## A. S. Bhattacharyya

Centre for Nanotechnology, Central University of Jharkhand, Ranchi: 835205

2006asb@gmail.com

**Abstract**

Coatings of ternary nanocomposite Si-C-N ceramic coatings have shown newer and improved mechanical and functional properties over the coarser and monolithic coatings. Properties like high hardness, wear resistance, oxidation resistance, tunable band gap and chemical inertness have been observed for Si-C-N which makes its potential for numerous applications. Although lot of research has taken place in Si-C-N coatings, proper understanding of the effect of different parameters on the coating properties are still not resolved. The changes occurring in fraction of Si, C and N and the phases forming in the coatings with variation in deposition conditions require investigations. This research paper gives a systematic study of the role of different deposition parameters like substrate temperature, pressure, power on the nucleation and growth, structure, microstructural bonding and mechanical properties of the film deposited by magnetron sputtering which adds significantly to the fundamental knowledge of nanocomposite Si-C-N coatings as well as its applications.

## 1. Introduction

The mechanical systems of today are exposed to complex operating conditions; therefore there is a demand for improvement in performance durability and efficiency of the components. Coatings of hard materials meet these demands by preventing wear and abrasion of the mechanical parts [1]. Nanoscale dispersion of fillers or controlled nanostructure in the composite [Fig-1] can introduce new physical properties effectively changing the nature of the original matrix which may not be possible in single conventional microstructure or higher constituent matrix [2-5]. It tends to drastically add to the electrical and thermal conductivity and also results in enhanced optical properties, dielectric properties or mechanical properties such as stiffness and strength[6].

In spite of a variety of materials with a high hardness, the coatings meeting all the requirements of high stiffness (high E) and sufficient stability against environmental attack is limited. Monolithic coatings of transition metal carbides/nitrides and oxides are mostly used as hard protective coatings. But conventional hard ceramic coatings like SiC, $Si_3N_4$, TiC, TiN, cBN diamond coatings fail to provide the optimum properties and at the same time expensive. Hard coatings of new compositions are always in demand if they could provide high performance and at the same time are less expensive than the existing. The new material is also required to be tougher. The relationship between hardness and wear has caused the scientific and engineering community to pursue the development of increasingly harder coatings but having reasonable toughness. Attentions however have recently been focused on duplex, functionally gradient multilayer and nanocomposite coatings. They show higher hardness than the presently available coatings. Hence new multicomponent nanocomposites are being studied globally [7-33].

Coatings of ternary nanocomposite Si-C-N ceramic coatings have shown newer and improved mechanical and functional properties over the coarser and monolithic coatings. Properties like high hardness, wear resistance, oxidation resistance, tunable band gap and chemical inertness have been observed for Si-C-N which makes its potential for numerous applications. Though investigation in the field of Si-C-N coatings has geared up globally in last few years, still the understanding of the mechanism, proper investigation into the microstructure, phase and mechanical behaviour is in infancy state and needs further study for understanding of the enhanced properties and their applications. In this communication a review of the structural and mechanical properties of these novel multifunctional thin films of nanocomposite Si-C-N system is given.

## 2. Silicon-Carbon-Nitride (Si-C-N) in bulk form

The ternary nanocomposite material Si-C-N was first introduced to the scientific community as a high temperature oxidation resistant polymer derived ceramic (PDCs) [34, 35]. PDCs are a new class of ceramics that combine the functional properties of polymers with the mechanical and chemical durability of ceramics. Polymer derived ceramics (PDCs) have a polymer-like nanostructure and ceramic-like properties, e.g., creep and oxidation resistance [36-47]. Organometallic polymer or oligomer (polysilazane) consisting of molecular Si-and N- compounds are used for Si-C-N. Initially cross-linking of this oligomer takes place on pyrolysis leading to amorphous to crystalline Si-C-N nanocomposite [48-54]. Pyrolysis of a carbon-rich polysilazane under inert atmospheres leads to an amorphous ceramic consisting of a Si-C-N-(H) phase and excess carbon in a trigonally bonded phase due to free-carbon content [40,41]

There are two families of polymer-derived ceramics: silicon oxycarbide (SiOC) and silicon carbonitride (Si-C-N). Both of them show unusual resistance to crystallization at high temperatures, far greater than the constituents from which they are made. Si-C-N resists crystallization to a temperature ($1600^{o}$C) higher than either SiC or $Si_3N_4$. Even though the X-ray diffraction pattern from these materials is usually featureless, they are not strictly amorphous. They contain short-range structural features, which we call nanodomains. The nanodomains are the basis for the remarkable resistance of PDCs to crystallization even at ultrahigh temperatures. The crystallization gets retarded because of extremely low atomic mobility in polymer-derived Si-C-N [42, 55]. Thus Si-C-N is stable up to high temperature of $1500^{o}$C and more.

Apart from high temperature oxidation resistance, the hardness as well as high temperature stability of Si–N–C phase exceeds those of SiC and $Si_3N_4$. The electronic band gap of 2.5-3.8 eV allows us to consider this material as a wide band gap and dielectric material. More importantly stability of β-$C_3N_4$ with comparable diamond hardness has been achieved in ternary silicon carbonitride .These properties make silicon carbonitride a promising material for prospective applications such as structural ceramics, MEMS, hard protective coatings, and electronic materials [43-48].

Although Si-C-N is basically amorphous, evidence of crystallization by hot pressing, variation of nitrogen pressure and plasma CVD has been found and reported in the literature [56-59]. This amorphous Si–C–N material undergoes a transition into the thermodynamically stable crystalline phases, accompanied by shrinking and cracking by a preliminary phase separation into C-rich, SiC-rich, and $Si_3N_4$ rich amorphous domains [42, 54]. The local structure around silicon atoms depends on the nitrogen/silicon ratio, regardless of the carbon amount.

The polymer-derived Si-C-N ceramics has been reported to be composed of microcrystalline $Si_3N_4$ grains surrounded by nanosized SiC crystallites. The free carbon is always the first phase to nucleate, independent of the Si : C ratio as so-called basic structural units (BSU) followed by the crystallization of SiC, which originates upon diffusion controlled thermo chemical degradation of the remaining amorphous Si-C-N matrix[50].

Si-C-N is a solid solution of $SiN_x$ and $CN_x$ where C atoms substitutes only for Si sites and vice-versa. The structure of Si-C-N similar to $\alpha$-$Si_3N_4$ which is a hexagonal Unit cell built up by corner sharing of $SiN_4$ tertrahedra. The local structure of silicon atoms depends on the relative amount of nitrogen and silicon, regardless of the carbon amount. Silicon atoms are always four-fold or higher coordinated and surrounded by the maximum possible number of nitrogen atoms [34, 35].

Only a nitrogen/silicon ratio around 4/3 leads to the formation of $Si_3N_4$ structures, and thus, towards a separation of an amorphous $Si_3N_4$ rich phase. Otherwise mixed $Si(C,N)_4$ tetrahedra are preferred [34,35]

Three different C phases are observed, depending on the nitrogen/silicon ratio, regardless of the carbon amount. For a low nitrogen/silicon ratio, the C atoms are spread in the system and exist as mixed $C(Si,C)_3$ mixed planar structures. For a nitrogen/silicon ratio around 4/3, a separation of amorphous C-rich domains is observed. For a high nitrogen/silicon ratio, when all the silicon atoms are screened by the nitrogen ones, crystalline graphitic monoatomic layers surrounding the Si-N domains are formed within the material. Amorphous Si-C-N preserve a tetrahedral random network or possess a dominant $Si_3N_4$ type and/or SiC type of short-range order network .There are 12 Si and 16 N atoms per unit cell space group P31c [48].

Si-C-N being a ternary structure, solid solution of $Si_3N_4$ and $C_3N_4$, evidence of formation of $\alpha$-and $\beta$-$C_3N_4$ , $Si_3N_4$ and SiC has been found. Small crystallites have been found to have nucleated in the amorphous C-N matrix. The size of the crystallites has been found to be less than 1$\mu$m.The d-values have been found to match with $\beta$-$C_3N_4$. Crystallites of sizes about 10 nm have also been reported on addition of Silicon. The films have also shown better mechanical property. Thus incorporation Si has aided in crystallization. However excessive Silicon addition led to the formation of amorphous phase [34,35 , 46-49].

Phases of α-and β-$C_3N_4$ are isomorphic to α-and β- $Si_3N_4$ respectively with all Si atoms replaced by C atoms. The unit cell of α phase is twice as large as β-phase. They differ in stacking sequence of the layer subunits along the z-axis. α -phase Si-C-N is a ternary structure solid solution of $Si_3N_4$ and $C_3N_4$. has stacking sequence of planes ABCDABCD while the β-phase- has ABAB stacking sequence. SiC on the other hand has got ACBACB,ABAB.ACABA and ACBACB for its 3C, 2H,4H and 6H polytypes [48] .

The covalent feature of C-N bond as well as the ionic behaviour of the SiN bond are more pronounced as more C atoms are introduced by substituting Si atoms in the α-$C_3N_4$.The population of strong covalent C-N bond in the α-Si-C-N crystal plays an important role in determining the hardness and other mechanical properties of the material in Si-C-N as more and more C atoms are introduced, substituting the Si atoms there is reduction of both bond length and lattice constant, the bulk modulus increase and reaches a maximum value in the α-$C_3N_4$ structure. However in real experimental situation the successful synthesis of C-rich α-Si-C-N crystal is difficult to achieve due to various forms of C-bonding to N (single, double, triple bond) same reason for lack of success with $C_3N_4$ crystals [39, 42].

In the Si-C-N phase diagram all the phases exist on the tie line between "$Si_3N_4$" and "$C_3N_4$ " evidence of formation of $SiC_2N_4$ stable at 400°C and the other is $Si_2CN_4$ stable at 920°C has been reported in the literature (**Fig 2**) .For Si rich α-Si-C-N the majority of network is formed by Si-N .C atoms substituted for Si atoms tend to keep $sp^3$ bonding configuration when the C content is low[38,39,40].

When the carbon content gets increased over a certain value, the formation of C-N, C=N and C≡N takes place, leading to the formation of α-Si-C-N crystal. The available α-Si-C-N crystal so far has been found to be Si rich having the C content below a critical limit. The included Si in the deposit substitutes the carbon atoms in the CN compound and bonded to the nitrogen. The incorporation of Si in the CN deposit promotes the inclusion of nitrogen and leads to $C_3N_4$.

The α-Si$_2$CN$_4$ and α-SiC$_2$N$_4$ phases in the Si-C-N system were found to be more symmetric than others. They are called pure crystalline α-Si$_{3-n}$C$_n$N$_4$(n=0, 1, 2, 3) phase. In α-Si$_2$CN$_4$ 4C atoms are substituted for Si atoms on four different stacking layers. While in α-SiC$_2$N$_4$ 8 Si sites on 4 different stacking layers replaced by C atoms. These two structure exhibit Cc symmetry which is identical to P31c when primitive unit cell vectors are same and angle between a and b is 120°. The two above mentioned crystalline structures can be called 'quasi hexagonal" with angle for α-Si$_2$CN$_4$ 119.13° and 120.93° for α-SiC$_2$N$_4$. The bond length being 1.43-1.48 Å for C-N and 1.71 − 1.74 Å for Si-N [34, 35].

In α-Si-C-N structures the difference in charge transfer from cation C or Si to anion N leads to different bonding characteristics compared to α-C$_3$N$_4$ and α-Si$_3$N$_4$ systems. In α-Si$_2$CN structure N atoms show less anion like properties with less effective charge transfer and a shorter Si-N bond(1.66 to 1.74) compared to those in α-Si$_3$N$_4$.The enhanced bonding characteristics is also reflected in other structural parameters such as bulk modulus. The average C-N bond in α-Si$_2$CN$_4$ is longer than that in α-C$_3$N$_4$ structure indicating less covalent character of α-Si$_2$CN$_4$ compounds than α-C$_3$N$_4$ [44].

The electrical conductivity (dc) for Si-C-N at room temperature is $10^4 \Omega^{-1} cm^{-1}$, due to formation of sp$^2$ carbon atoms. The refractive index of oxidized Si-C-N is 1.4 and that of nonoxidized is 2.07. The band gap is adjustable between 2.86 to 5.0eV.The extinction coefficient is 10.0 for oxidized and 2.0 for nonoxidised. Fracture toughness of 2.1 MPa√m from indentation crack length measurement and 3.3-4.3 MPa√m from single edged notch beam method for have been reported. Better creep is also shown by these materials compared to monolithic ceramics. Better sinterability at lower temperature and easier machining of sintered products are also found. Improved thermomechanical properties after sintering compared to monolithic SiC and Si$_3$N$_4$ have been reported. Viscosities up to $10^{16}$ Pa as in the amorphous form has been measured. Si-C-N films can be deposited even on substrates without thermal stability, such as plastics. The Si-C-N has shown stability up to 1600°C and it remains in the amorphous state up to 1500$^o$C [34, 35 ,45-58].

The reason for these unique properties shown by Si-C-N is complex covalent bonding, cross-linking of Si, C and N atoms, low oxygen diffusion, small grain size, large volume fraction of the interface, compressive residual stress, inter grain amorphous layer , growth orientation and solid solubility of different phases. The high hardness shown is due to the formation of phases like SiC, $Si_3N_4$ and $C_3N_4$[34,35,52].

During Si-C-N film growth there is competition among various SiN, SiC and Si-C-N crystallites. Layers of Si-O, Si-C-N and Si-N layers are formed at different deposition times. Si-O and Si-C-N are formed initially whereas Si-N is formed after long deposition time. Si-C-N layer consists of nanosized crystallites embedded in the amorphous matrix .The crystallites are however close to the SiC crystal structure. SiN layers consist $Si_3N_4$ phases. The deposited films have been mainly found to be amorphous. Crystalline $\beta$-$C_3N_4$ found to be stable in the presence of $\beta$-$Si_3N_4$.Thus both these phases should mix to form $SiC_xN_y$ phase [34, 35, 52].

The dominant phases in the Si–N–C system are SiC and $Si_3N_4$. They are immiscible as the incorporation of carbon into the $Si_3N_4$ lattice is a difficult task. Si–N–C phases (amorphous for X-rays) have been prepared by the following synthesis methods, pyrolysis of polyorganosilazanes (POPOS), chemical vapor deposition (CVD), and physical vapor deposition (PVD) and this immiscibility of SiC and $Si_3N_4$ has been overcome by using an organosilazane precursor with existing bonding between Si, N, and C atoms. Crystalline Si–N–C phases have been reported to be grown by microwave plasma CVD [59].

## 2. Applications of Si-C-N

Si-C-N nanocomposites due to their exceptional combination of properties mentioned above are very promising material for the future. They have been proved ideal for applications as wear and oxidation resistance, optoelectronic material in hostile environment [34, 35, 37]. Si-C-N coatings are desired for many industrial applications such as turbine engines, blades and wear resistant coatings for automotive industry to enhance the life as well as the performances of the components. They also find application in MEMS device fabrication in the form of field emission displays, catalyst support, high temperature semiconducting device, metal and polymer matrix composites and high temperature applications[43,44,60-63].Smooth surface morphology, stable amorphous structure good prospect electronic devices [43, 44].

Si-C-N being a wide band gap material and having capability of tunable band gap has the potential application in optoelectronics. It has got good optical transmittance properties for visible and IR region. Due to low density and good thermal shock resistance are very important requirement for future aerospace and other wear, oxidation resistance applications [52]. Polymer-derived Si-C-N ceramic has been used for making micro-gear of 200 mm diameter and 40 mm thickness, fabricated with the micro-casting technique. They have been used for nanomechanical resonators. They have also been used for making of microigniters [44].

More recently Si-C-N with large (several tens of microns) and well-facet crystals have been grown successfully by microwave plasma-enhanced chemical vapor deposition (MW-CVD). These newly discovered Si-C-N crystals have been found to demonstrate a great potential for applications in semiconductor technology as well as a structural material. They show excellent mechanical properties with hardness and bulk modulus around 30 and 322 GPa, respectively. The optical investigation of polycrystalline Si-C-N films shows a direct band gap of about 3.8 eV and a strong optical emission around 2.8 eV. The ternary Si-C-N compound constitutes an important wide band gap material with gap energy within the blue-UV spectral region.

### 3. Si-C-N Coatings

The material Si-C-N has been deposited as coatings on various substrates by different methods [64-92]. Si-C-N coatings provide resistance to wear, abrasion, corrosion and oxidation. It has high thermal conductivity, thermal stability (upto1500$^{o}$C), oxidation stability[82], high hardness[78], wide band gap [59,74], chemical inertness, promising wetting behaviour, wear resistance, excellent chemical stability, and promising mechanical, thermal ,electrical [81]and optical properties[74].

Several methods for the fabrication of amorphous and crystalline Si-C-N films are reported in literature. Both crystalline and amorphous or nanostructured Si-C-N compounds have been prepared. Thin film depositions of Si-C-N have been carried out by Plasma and ion assisted deposition [64, 65, 66], Pulsed high energy plasma deposition [67], Chemical Vapour Deposition [64], sputtering [85,91,92], microwave and electron cyclotron resonance PECVD [88,90], ion implantation [92] , pulsed laser deposition [84],

HWCVD [71], electrochemical methods [70], remote plasma CVD [76] and Rapid thermal Chemical vapour deposition. Nanocrystalline $Si_2CN_4$ has been prepared by metal organic CVD.

Although the deposited Si-C-N are mostly amorphous, nanocrystalline $Si_2CN_4$ has been successfully prepared through metal organic CVD. Growth and mechanical properties of large (several tens of microns), well-faceted ternary Si-C-N crystals grown by microwave plasma-enhanced chemical vapor deposition and amorphous Si-C-N films grown by ECR-CVD [88, 90] have also been reported.

Other techniques used to synthesize Si-C-N alloys are reactive sputtering [81] and ion assisted dynamic mixing. Carbon–silicon–nitride thin films were grown on (10 0) oriented silicon using pulsed laser deposition (PLD) assisted by a RF nitrogen plasma source It was found that due to incorporation of 10% $Si_3N_4$ to the PLD graphite target the $CSi_xN_y$ films show a hardness value of 23 GPa (reference value for silicon substrate 14 GPa), but strongly decreased elastic modulus from 464 to 229 GPa. Amorphous Si-C-N films synthesized by thermal chemical vapor deposition (CVD) techniques showed a hardness in the range of 27–38 GPa and a stiffness of 224–289 GPa whereas a-Si-C-N films deposited by electron cyclotron resonance plasma CVD possessed a hardness of 22 GPa[88,90]

Amorphous $SiC_xN_y$ films deposited with various compositions by ion beam sputtering showed decrease in Youngs modulus with increase in carbon content [92]. The hardness and effective modulus of hydrogen-containing and hydrogen-free amorphous $CSi_xN_y$ films, deposited by electron cyclotron resonance plasma chemical vapor deposition (ECR-CVD) and hydrogen free ion-beam sputtering deposition (IBSD), respectively, were studied through nano-indentation [88,90,92]. Amorphous Si-C-N films were deposited on Si (1 0 0) substrate by nitrogen ion-assisted pulsed-laser ablation of an SiC target and the dependence of the formed chemical bonds in the films on nitrogen ion energy and substrate temperature were investigated by XPS [84].

Hence it can be seen that although there are reports on the deposition of Si-C-N by sputtering [85, 91, 92], not much has been discussed on the effect of all the plasma parameters; which justifies the novelty of the work reported.

### 4. Si-C-N coatings by Magnetron sputtering

An alternative way of mixing the three atomic components is by *magnetron sputtering* using SiC targets in argon/nitrogen plasma. This method allows one to produce Si–N–C amorphous films. Sputtering is a method of depositing both thin metal films and insulators onto a substrate [8, 94, 95, 96]. Unlike evaporation, the material to be sputtered does not have to be heated. The deposition of alloys and insulators as composite materials are two important benefits of sputtering [8, 94, 95].

In magnetron sputtering the ejection of the atoms from the surface of material (The target) is carried out by bombardment with energetic particles in the presence of magnetic field which enhances the plasma density due to recoil motion of the electrons perpendicular to both electric and magnetic field. Plasma is a fluid of positive ions and electrons in a quasineutral electrical state. In many plasma coating applications, positive ions are generated by collisions between neutral particles and energetic electrons. The electrons in plasma are highly mobile, especially compared to the larger ions (Usually argon ions for sputtering). Control of these highly mobile plasma electrons is the key to all forms of plasma control. The ions are accelerated by the positive ion sheath at the cathode and cause sputter erosion of the cathode material. There is substantial induced current, which flows parallel to the cathode surface and is located adjacent to the cathode ion sheath. This self closing flow of electron is along the sputter erosion track and rurally called the $E \times B$ drift current. However only large electric field in a magnetron discharge occurs in the cathode ion sheath which is typically very thin and does not encompass or directly cause the drift current. With the help of magnetron the plasma can be sustained even an at lower pressure which leads to cleaner and better adhering coatings, also due to ion density the deposition rates are higher In magnetron sputtering, substrate heating is eliminated due to lesser bombardment of secondary electrons. Among the various deposition techniques sputtering shows excellent adhesion and high deposition rates. However some disadvantages of sputtering are low target material utilization, high deposition rates that are sometimes undesired [8, 94, 95].

The sputtering process takes place in an evacuated chamber. Argon is introduced, which gets ionized in the chamber containing the substrate and the target of the film material to be sputtered. The ionized argon atoms impinge the target leading to a momentum transfer to the target atoms from the impinging ion [94, 97, 98]. The target is maintained at a negative potential relative to the positively charged argon atom. The positive ion accelerates towards the negative charge, striking the target with sufficient force to remove material. The argon atom does not become imbedded in the target. It slams into it like a steel ball into the wall and tears off some of the target material. Since the chamber is maintained at a vacuum, the liberated material settles on everything in the chamber, mainly the substrates (*Fig 2.5*).

In diode sputtering, not all of the electrons escaping the target contribute to the ionized plasma glow area. The wasted electrons fly around the chamber causing radiation and other problems, for example, the heating of the target and also lower deposition rate [8, 94, 99-102]. A magnetron sputtering source addresses the electron problem by placing magnets behind, and sometimes, at the sides of the target. These magnets capture the escaping electrons and confine them to the immediate vicinity of the target. The ion current (density of ionized argon atoms hitting the target) is increased by an order of magnitude over conventional diode sputtering systems, resulting in faster deposition rates at lower pressure [8,103-108]. The lower pressure in the chamber helps create a cleaner film. Target temperature is lower with magnetron sputtering enhancing the deposition of high quality films [8, 94, 95]

Thin pure homogenous uniform thin films are in demand today to obtain new and controlled effects in optics and electronics. High quality vacuum is required for the deposition of such films as atmospheric gases may cause contamination. Magnetron sputtering involves the application of both electric and magnetic fields to the conventional DC diode and sputtering cathode. The electric field can be either DC or RF. High deposition rates are obtained through this method due to cross-field patterns [8]. The advantage that magnetron sputtering has over conventional sputtering is that the substrate heating is avoided the latter case. In conventional sputtering however the substrate sputtering takes place due to bombardment of secondary electrons. The secondary electrons in magnetron sputtering are not allowed to reach the substrate. Hence very little heating takes place, thus heat sensitive substrates, materials can be deposited. This method is also comparatively much cleaner and leads to homogenous film depositions. the deposition rate is also higher as the method is more energetic and phase formation can take place at lower temperature [94,95].

Silicon carbonitride (Si-C-N) films were deposited by radio-frequency reactive sputtering and then annealed at 750 °C in nitrogen atmosphere [89, 91]. The $N_2$ fraction in the gas mixture has been reported a major influence on the deposition rate and composition of the films deposited by magnetron sputtering: A high $N_2$ fraction results in increased deposition rate and N/(C + Si) and C/(N + Si) ratios. The Si-C-N films are mainly amorphous. XPS and FTIR spectra show the existence of Si–N, C–N single, C=N double, and C≡N triple bonds. Almost no Si–C bonding is found in the films.

Increase in methane/nitrogen ratio led to a change from mirror-like smooth films to column-like and ridge-like C-rich $SiC_xN_y$ nanostructures. Micro-Raman studies have shown blue shift and narrowing of the G band at higher methane concentrations, suggesting an increase in the short-range order of the graphite-like phase in the nanostructured films [105]. The sharper geometric features of the nanostructured $SiC_xN_y$ films and the higher conductivity of the films led to an enhancement in field emission properties. A low turn-on field (<10 V $\mu m^{-1}$) and high emission current density (>0.2 mA $cm^{-2}$), as well as good temporal emission stability, have been achieved for the magnetron sputters nanostructured $SiC_xN_y$ films [70, 85].

The effect of $N_2$ in terms of variation of pressure has also been reported in our earlier publication.

Si-C-N films have been produced along the tie line $SiC–Si_3N_4$, $SiC–C_3N_4$ and $Si_3N_4–C_3N_4$ by means of reactive magnetron sputtering of a silicon target in an Argon/Nitrogen/Acetylene atmosphere. Depending on their position in the Si–C–N phase diagram, the hardness of the films varies over a broad range, with maximum values of around 30 GPa, while Young's modulus remains in a narrow range around 200 GPa [105,108]. XPS spectra showed the main component to be Si–C but Si–N and to a lesser extent C–C bonds were also detected. Further, IR spectra suggested the presence of the carbodimide group. Raman Spectra show a varying ratio of $sp^3$ to $sp^2$ carbon, depending on deposition condition [106]. The hardest films were found along the $SiC–Si_3N_4$ tie line. These films exhibited the highest load carrying capacity in lubricated sliding as well as the best adhesion [107]. In dry sliding their brittleness coupled with a high coefficient of friction led to premature coating failure. Carbon rich films have a very low coefficient of friction leading to good wear behaviour in dry conditions, but their ability to withstand high pressures is reduced [90]. The low coefficient of friction is attributed the more graphitic structure of the free carbon in the films. Good wear properties in dry reciprocating sliding as well as high load carrying capacity in lubricated sliding were achieved by producing gradient films from silicon rich to carbon rich at the surface [107].

Silicon carbon nitride thin films were also prepared by microwave ECR plasma enhanced unbalanced magnetron sputtering [93]. The hardness of the thin films studied by nano-indentation, were found to increase with the graphite target voltage and reached a maximum value of 25 GPa [93].

Si–C–N films were deposited on stainless substrates with different nitrogen flow rate using magnetron sputtering have shown formation of Si-N, C≡N, nitrogen-bound $sp^2$- and $sp^3$- hybridized carbon bonds. In the Raman spectra, the position of the D band was found to decrease with increasing nitrogen flow rate while the FWHM and the relative intensity of the D band to G band first decrease and then increase with increasing nitrogen flow rate. This indicates that the D band is not only related to the nitrogen contents, but also to the chemical environments or bonding nature of the C and N atoms. It may be closely correlated with the formation of $sp^3$ clusters. XPS measurements confirmed the presence of the bonds revealed by Raman spectroscopy at various nitrogen flow rates. These imply that the physical features of Si–C–N films can be well controlled or tailored by changing the nitrogen concentration [109].

The Si-C-N films prepared by reactive magnetron sputtering using sintered SiC target was found to posses a very complex network structure rather than a simple mixture of SiC, Si N and CN. Both FTIR and XPS analysis revealed the presence of chemical bonding among Si, N, and C components. The electronegativity of the third element played an important role in the chemical shifting of the binding energy between the other two elements. Quantitative analysis showed that the composition of as-deposited film was very close to the stoichiometry of SiCN. Furthermore, it has a smooth surface morphology and a stable amorphous structure, which offers good prospect for application in electronic devices [91].

Amorphous Si–C–N films were deposited on p-type Si(100) substrates by dc magnetron co-sputtering of silicon and carbon using a single sputter target with variable SiC area ratios in nitrogen–argon mixtures. It was shown that with a rising argon concentration in the gas mixture, the Si content in the films rapidly increases Si fraction in the erosion target area while the C content decreases at an almost constant N concentration. An intensified bombardment of growing films by argon leads to its subplantation into the films and to a decrease in a volume concentration of hydrogen .As a result, the N–Si and Si–N bonds dominate over the respective N–C and Si–O bonds, preferred in a pure nitrogen discharge, and the film hardness increases up to 40GPa. A decrease in the substrate temperature values leads to a higher incorporation of hydrogen into the films and to a stronger influence of the negative substrate bias voltage [110].

Increasing the silicon fraction in the erosion target area at a fixed Ar gas mixture resulted in a rapid rise in optical transparency of the films and in their friction coefficient (from 0.24 to 0.53) at almost constant film hardness close to 30GPa. Increasing the argon concentration in the gas mixture from Si fraction resulted in rise in the film hardness, while the stabilized values of the friction coefficient are practically constant (0.40– 0.46) in spite of the significant changes in the film composition and in the surface bonding structure. The increasing hardness was caused by direct subplantation of argon into the films and by ion-induced knock-on subplantation of the film forming particles, which lead to densification of the films [111].

SiCN$_x$ coatings have been also prepared by magnetron sputtering of a carbon target in a N$_2$ atmosphere adding a vapour pressure of Si(CH$_3$)$_3$Cl on high speed steel substrates. The presence of silicon Si allows the maintenance of a low friction coefficient independent of the relative humidity with a wear resistance comparable to pure CN$x$ coatings. The better tribological behaviour of SiCN$_x$ films under humid atmosphere was explained as being due to the adsorption of water molecules on SiO$_2$ domains present in the films. The water molecules may lubricate the contact allowing the decrease of the shear strength [112].

The methane fraction in the process gas for reactive magnetron sputtering and the substrate temperature were found to have strong effects on the film composition and morphology, as well as on the field emission characteristics. The etching effect of methane is responsible for the nanostructure morphology of the SiC$_x$N$_y$ films deposited at high temperatures. Carbon-rich a-SiC$_x$N$_y$ films with ridge-like microstructures were observed for films prepared with high methane fractions and high substrate temperatures. Furthermore, high methane fraction in the process gas also produced films with enhanced short-range order of the graphite-like phase and increased conductivity [85].

SiCN films were prepared by reactive magnetron co-sputtering from Si and C targets. The microstructure was found to predominantly amorphous-to-graphite like for the ternary Si-C-N system with low silicon concentrations at substrate temperature of 350$^o$C. For the higher Si contents, films showed a nanocomposite microstructure consisting of nanocrystals in an amorphous matrix. For low Si concentration, carbon atoms bond to carbon and nitrogen, and silicon atoms bond to nitrogen. At a high Si concentration, mainly Si-N, Si-C and C-C phases were present and less C-N bonds are observed. The Si-C-N films showed comparable, but slightly lower hardness values to what has been reported for CN$_x$ films grown under similar conditions, while the elasticity of the films showed to be not as high as for CN$_x$. The highest hardness value for a 1µm-thick film grown in this study reached 28GPa. [113]

## 5. Structure - property correlation of magnetron sputtered Si-C-N coatings

Investigation of mechanical behaviour of magnetron sputtered Si-C-N film deposited on silicon, glass and steel substrates at various deposition parameters have been investigated. The variation of particle sizes and the nucleation and growth of different phases at different deposition conditions were observed.

The x-ray diffraction (XRD) analysis of the SiC target *(Fig.2a)* used as the sputtering source showed the presence of *β-SiC* and C phases on the diffraction pattern [114]. However the XRD patterns of all the films deposited at different pressures showed amorphous nature. Even if a small amount of crystallization had taken place during deposition of these films, the XRD was not able to show any characteristic peak in the diffraction pattern and due to small volume percentage, only peak corresponding to the substrate was visible [115, 116]. This is a common phenomenon for the crystallization of Si–C–N materials, which has been observed by several groups [117].

The depositions of the Si-C-N coatings were carried out at four different pressures of 0.4, 1, 5 and 10 Pa. Higher deposition rate was observed in the range 1Pa-5Pa compared to the other higher and lower pressures, adjustments in deposition time were done in order to get coatings of similar thickness. The thickness measurement was performed from a step obtained using surface profilometer at the interface between the coated and uncoated substrate. A typical thickness profile is shown in *Fig 2 (b)*. The step height was the thickness of the coating. Lower deposition rate at higher and lower pressure was due to more ion-electron collisions at higher pressure that had neutralized as mean free path was smaller and at lower pressure, the mean free path was larger which led to less ionization of the plasma and hence less removal of target atoms. The deposition rate on the other hand due to resputtering from the substrate, was found to decrease with increase in temperature. However, in the case of power variation, the deposition rate increased with increase in power as the intensity was more.

The average surface roughness was measured using surface profilometer for a large area of the coating. An average roughness in the range of 10nm was found. A typical roughness profile showing is given in *Fig 2(c)*.

The effect of nitrogen was found to be significant while investigating the effect of pressure using AFM [115,116]. Although an initial increase in nitrogen pressure helped in formation of $Si_3N_4$ and $C_3N_4$, its higher percentage led to formation of graphitic phase due to nitrogen desorption. A certain range of argon-nitrogen pressure in the range 1-5 Pa was identified to be giving films with smaller particle size, lower roughness and also higher hardness as reported by S.K.Mishra et al.

A set of plots of hardness and modulus obtained from nanoindentation of Si-C-N on steel substrate at different pressure along with the hardness and modulus plot of the bare 304SS substrate is shown in Fig 3 (a) and (b). The hardness values given are the average of four nanoindentation experiments. All the plots show an initial increase in H and E values with depth which is due to fully elastic response of the coating towards indenter penetration. The response changed from elastic to elastic-plastic and lastly to fully plastic as the load and correspondingly the depth of penetration were increased. The maximum values H and E obtained during their variation with depth of penetration, are the hardness and modulus of the film. The decrease in H and E values after the highest values were achieved was due to the effect of comparatively softer steel substrate having a hardness of 3.5 GPa and corresponding modulus of 210 GPa.

An increase in H and E was observed on increasing the pressure from 0.4Pa to 1Pa. Coatings deposited in the range of 1Pa $Ar/N_2$ pressure showed higher hardness and modulus of 25GPa and 250GPa than 0.4 Pa. However, a decrease in the values were observed on increasing the pressure further to 5Pa and lowest hardness of 2 GPa and modulus of 80 GPa were observed for coatings deposited at 10 GPa.

The load depth curve for the highest obtained hardness value was compared with that of the bare steel substrates as shown in Fig 4(c). Although, the maximum displacement was initialized to 500 nm for both the cases, the indenter had to apply a load of 110 mN to achieve that depth in case of the coated sample compared to a load of 25 mN for the bare substrate. The slope of the unloading curve is also more for the coating indicating higher stiffness than the substrate. The area under the load depth curve is the plastic energy which is higher in case of the bare steel substrate indicating higher plastic deformation to be taking place and consequently lower hardness.

Similar trend in hardness and modulus variation with depositin pressure was obtained for Si-C-N coatings deposited on glass and silicon substrates (Fig 4 a-d)

A detailed investigation of the effect of substrate temperature on the structural and mechanical properties of Si-C-N coatings using TEM and nanoindentation studies, has been reported by A.S. Bhattacharyya et al [118]. From TEM studies, it was observed that an initial increase in substrate temperature led to a transition from nc-graphitic phase to amorphous carbon which reacted with Si and N and eventually formed the amorphous Si-C-N matrix. At higher temperatures, nucleation of nanosized crystallites of hard phases of $\beta$-$Si_3N_4$ and $\beta$-$C_3N_4$ occurred in the amorphous Si-C-N matrix. The crystallite size varied between 2 to 20 nm on varying the substrate temperature from RT to 500$^o$C. However, on further rise in temperature to 600$^o$C, increase in the growth of the carbon phase due to desorption of nitrogen at higher temperatures occurred, which led to a decrease in hardness as reported in ref [118]. A representative set of micrographs of Si-C-N coatings at RT, 100$^o$C, 500$^o$C and 600$^o$C are shown in Fig 5. The E values of the coatings follow the same trend as hardness and shown in Fig 6(a, b and c) for Si, Glass and steel substrates.

The Raman spectra of Si-C-N coatings deposited at different substrate temperatures are shown in Fig 7. The spectra have been deconvoluted and discussed in details in our communicate manuscript to Journal of Raman spectroscopy [119]. The transition from one carbon form to another leading to the variation in hardness and modulus is due to the variation in $sp^2/sp^3$ carbon percentage and results in shift in G (graphitic) and D(disorder) peak position and intensity.

In the spectra for RT deposition, a G band position at 1571cm$^{-1}$ and D band position at 1350 cm$^{-1}$ are observed which corresponds to nc-graphite. On increasing the temperature further, the crystallite size continues to lower further which results in disappearance of nc-graphitic phase and formation of $a$-Si-C-N. The prominent separation between the D and G positions, which is the evidence of graphitization, gradually decreases with increase in temperature and almost a single band formation at 1500 cm$^{-1}$ occurs at 400$^{o}$C. This corresponds to the formation $sp^3$ hybridized $ta$-C in the form of nucleation and growth of $\beta$-$C_3N_4$ crystallites in the $a$-Si-C-N matrix. This results in an upper shift of the G band position as observed in *the spectrum* for coating deposited at 500$^{o}$C. The microstructural and structural changes have been discussed above on increasing the substrate temperature. At 600$^{o}$C however, a further increase in the G band position observed. This is due to the growth of carbon crystallites and formation of nc-graphitic phase [119].

An increase in power from 200W up to 400W keeping the pressure and temperature at 1Pa and 500$^{o}$C, leads to better coalescence of grains resulting in increased hardness and modulus for all coatings deposited on the substrates *(Fig 8 a-d)*.

The highest hardness of 4825 Hv (Vickers microindentation) and 25GPa (nanoindentation) with elastic modulus of 250GPa were obtained in case of Si-C-N coatings deposited on 304SS substrates at 500$^{o}$C substrate temperature, 1Pa pressure and 400W power [120]. In the case of silicon substrates, for the same deposition parameters, the values were 4445 Hv (Vickers microindentation), and 22 GPa from Nanoindentation with bulk modulus of 240 GPa. In case of glass substrates the values were 4305 Hv from microindentation 17GPa from nanoindentation and 150 GPa elastic modulus. A correlation of the structure and hardness of rf magnetron sputterd SiCN coatings using FTIR and nanoindentation has been published elsewhere [121].

The highest obtained hardness and modulus in a nanoindentation profile values are taken as the hardness of the coating for hard coatings on comparatively softer substrate. The variation of modulus and hardness values with variation in plasma parameters for all the three substrates are shown in Figs 9, 10, 11.

The hardness value obtained by nanoindentation showed lower values than microhardness due to different mode of hardness measurements. In microhardness, it is measured by measuring the diagonal lengths ($d_1.d_2$) of the plastically deformed square indentation impression using a four sided pyramidal Vickers indenter after the recovery of the surrounding elastic matrix as shown in *Fig 12(a, b and c)*. Hardness as well as modulus estimation in case of nanoindentaion on the other hand, unlike measurements related to permanent plastically deformed region in microindentation, are determined through a continuous loading-unloading process (*Fig 12 (d)*) which nullifies the effect of elastic recoveries. Also the use of nanometric highly machined three sided Berkovich nanoindenter leads to comparatively higher plastic deformation than the Vickers microindenter (*Fig 12 e and f*). As the hardness will be more where the elastic recovery is more and plastic deformation is less, the microhardness values were found more than the nanohardness values. Another observation that can be made by looking at the nanohardness and modulus curves was the effect of substrate leading to lowering of values starting earlier in case of modulus than in hardness. The reason being: H values are obtained from the loading process whereas E is obtained from the unloading process and the effect of substrates is earlier on the unloading process.

Comparing Si-C-N coating deposited in RF and DC modes, it was observed that the films deposited in RF mode were harder and tougher compared to the DC deposited films. The comparative study of the structure and properties of Si-C-N films deposited in RF and DC has been reported by A.S.Bhattacharyya et al[122].

Adhesion studies using scratch tests showed a strong adherence between the coating and the substrate as reported by S.K.Mishra et al[123]. The scratch tests were performed at a ramping load and the point of coating failure called the critical load was considered as the measure of adhesion. The highest critical load of 24 N was obtained for coatings deposited on 304SS substrates. Combined affect of crystallization and residual stress were observed on the adhesion of the Si-C-N coating. The failure modes in scratch tests also had a major effect from the substrate. The films deposited on silicon substrates showed brittle nature of failure which consisted buckling and radial crack formation. The coatings deposited on glass substrates showed wedge spallation and Hertzian circular cracks. Plastic deformations consisting of pile up on the sides of the scratch track were found for coatings deposited on steel substrates [123,124].

Thus the nanocomposite Si-C-N coating properties were found to be strongly dependent upon the deposition parameters viz., $N_2$-Ar pressure maintained in the vacuum chamber, substrate temperature, power applied and mode of deposition (RF and DC). The different ratios of phases of $\beta$-$C_3N_4$, $Si_3N_4$, graphite in the film and the effective particle size at different deposition conditions had an effect on the hardness of the films. Hardness was found to be highest for Si-C-N coatings deposited at 1Pa pressure, 500$^\mathrm{o}$C substrate temperature and 400W. XPS of SiCN coatings are published in ref [125, 126]. Fractographic studies and other mechanical aspects including scratch tests. Computational nanoindentation studies and substrate effect are given in ref [126-130].

**Acknowledgements**

**Legends to figures:**

1. (a) Nanocrystalline fillers in amorphous matrix forming nanocomposite (b) Ternary phase diagram of Si-C-N.

2. (a) XRD of the target (b) Thickness profile of Si-C-N film on Silicon measured from surface profilometer (c) Average roughness of 10nm from surface profilometer.

3. (a) Hardness and (b) Modulus obtained through nanoindentation of SiCN coating on 304SS at different pressure and corresponding (c) load-depth curve for the bare 304SS substrate and coating deposited at 1 Pa[118-126].

4. Hardness and Modulus obtained through nanoindentation of SiCN coating on (a, b) silicon and (c,d) glass at different pressures[118-126]..

5. TEM of Si-C-N coating deposited at (a) RT (b) 100$^o$C (c) 500$^o$C and (d) 600$^o$C [118].

6. Elastic modulus plots of SiCN coatings deposited on (a) silicon (b) 304SS and (c) glass substrates at different substrate temperatures[118-126].

7: Raman spectra of Si-C-N coating deposited at various substrate temperatures [119].

8. Hardness and elastic modulus plots obtained through nanoindentation of Si-C-N coatings deposited at different power on (a, b) 304 SS and (c, d) glass substrates [118-126].

9. Variation of hardness and modulus obtained through nanoindentation for coatings deposited on (b) silicon, (c) glass and (d) steel with pressure [118-126]

10. Variation of hardness and modulus obtained through nanoindentation for coatings deposited on (b) silicon (c) glass and (d) steel with temperature [118-126].

11. Variation of hardness and modulus obtained through nanoindentation for coatings deposited on (b) silicon, (c) glass and (d) steel with substrate temperature [118-126].

12. (a) Measurement of hardness through measurement of diagonals and (b) top view and (c) cross-sectional view of elastic recovery in Vickers microindentation(d) Measurement of H & E through continuous loading-unloading process in nanoindentation (e) Four sided Vickers indenter and (f) three sided sharper Berkovich indenter [129].

Figure-1

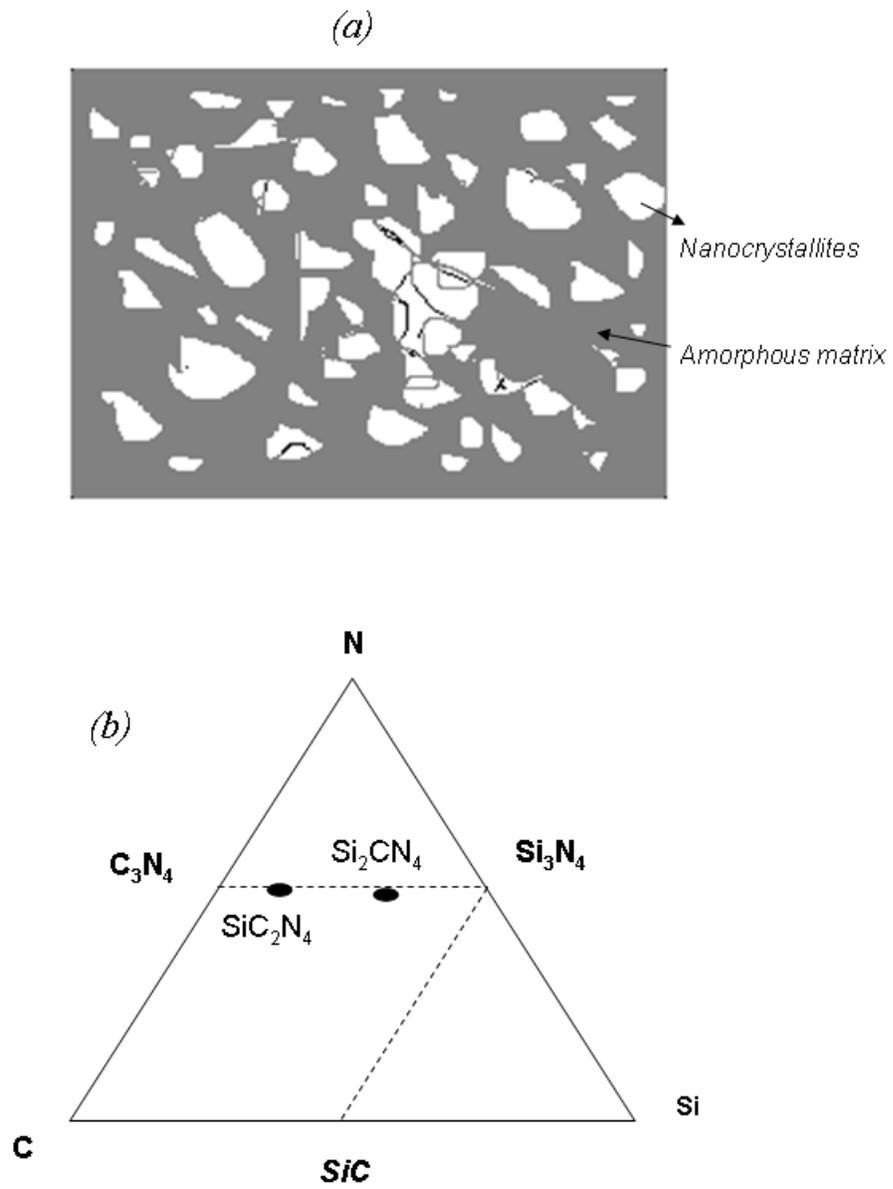

*(a)*

Nanocrystallites

Amorphous matrix

*(b)*

N

C₃N₄  Si₂CN₄  Si₃N₄

SiC₂N₄

C  SiC  Si

Figure-2

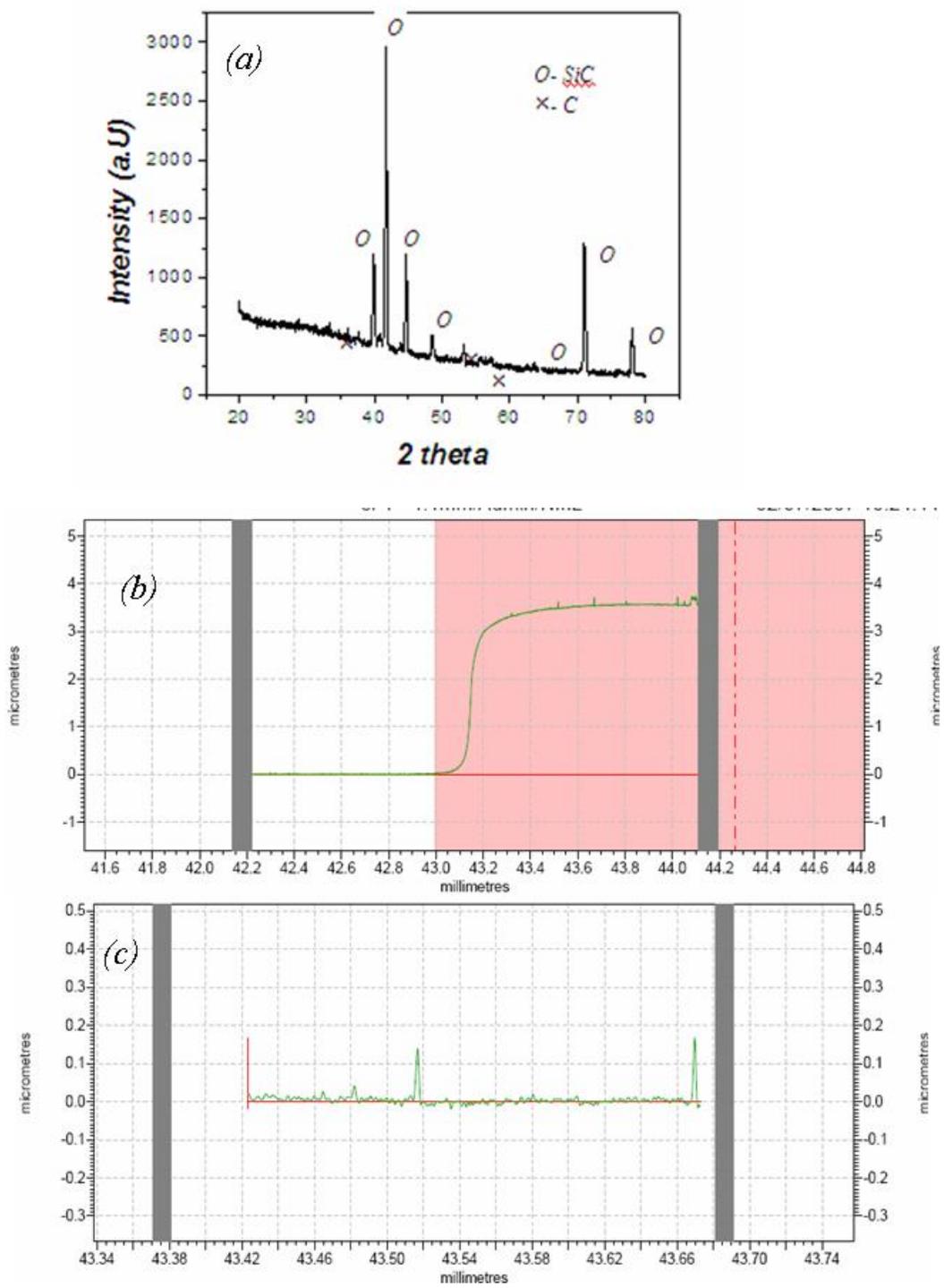

Figure-3

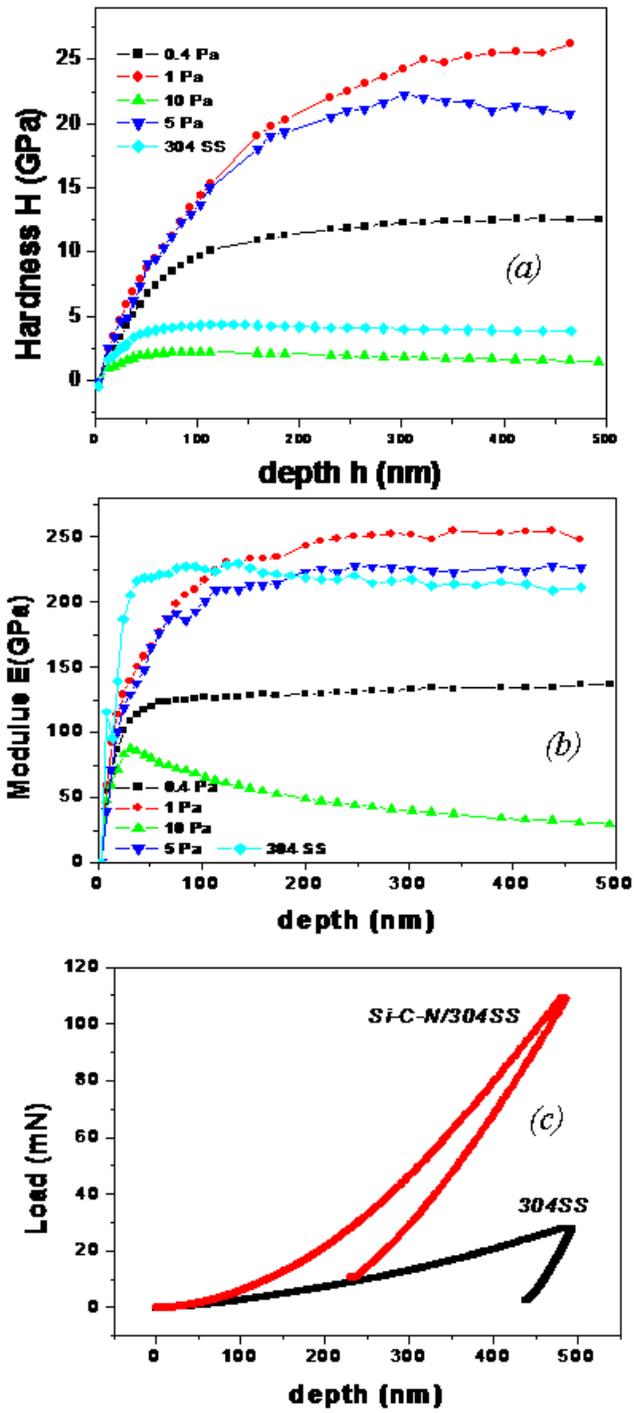

Figure-4

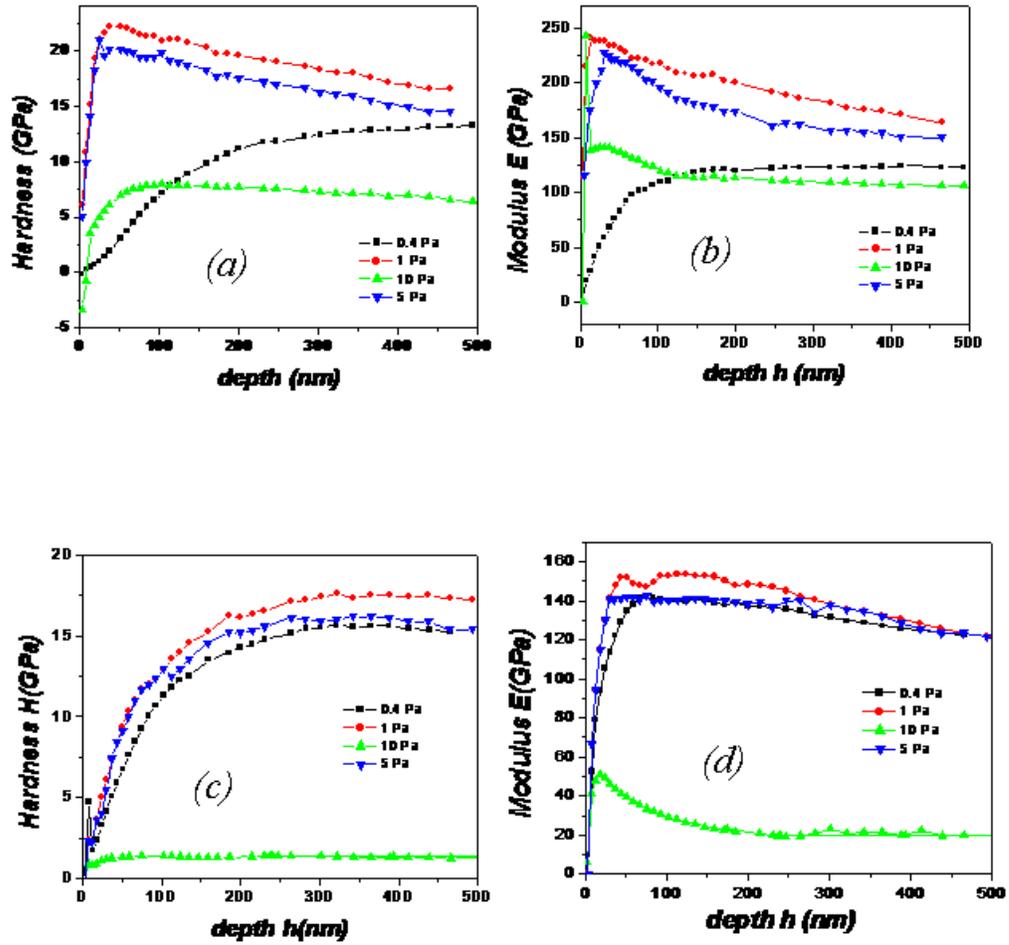

Figure-5

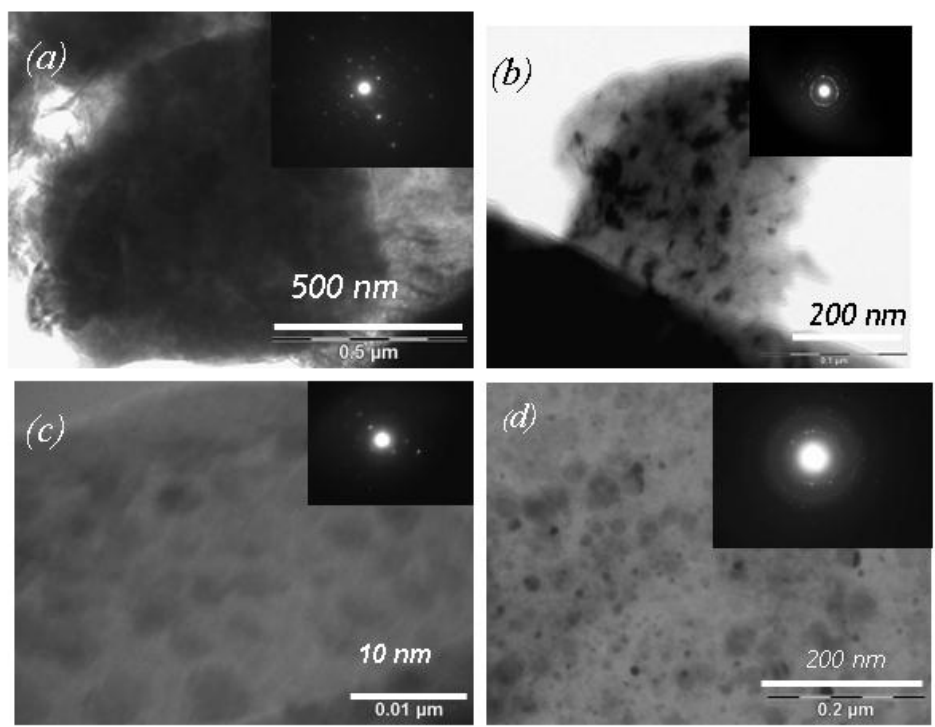

Figure-6

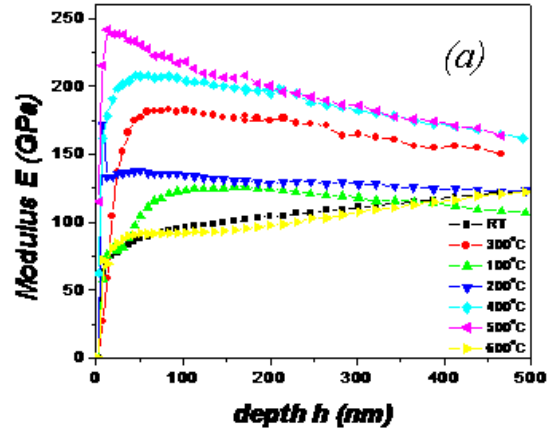

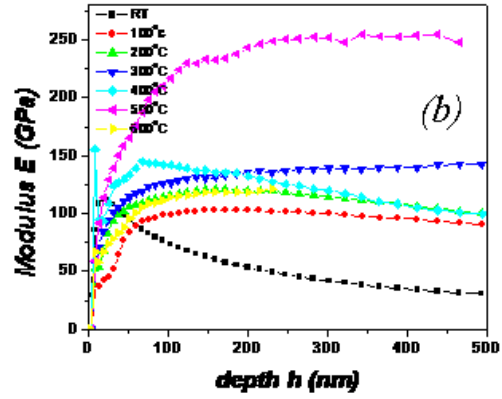

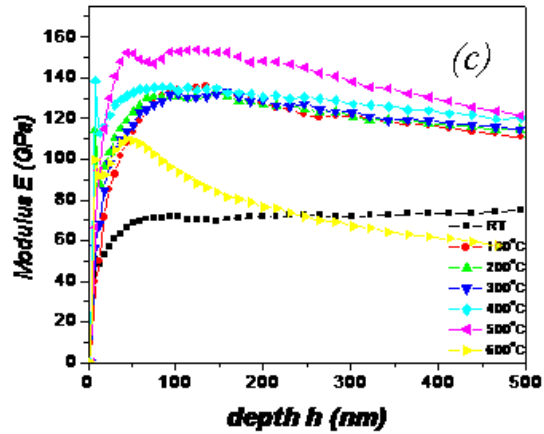

Figure-7

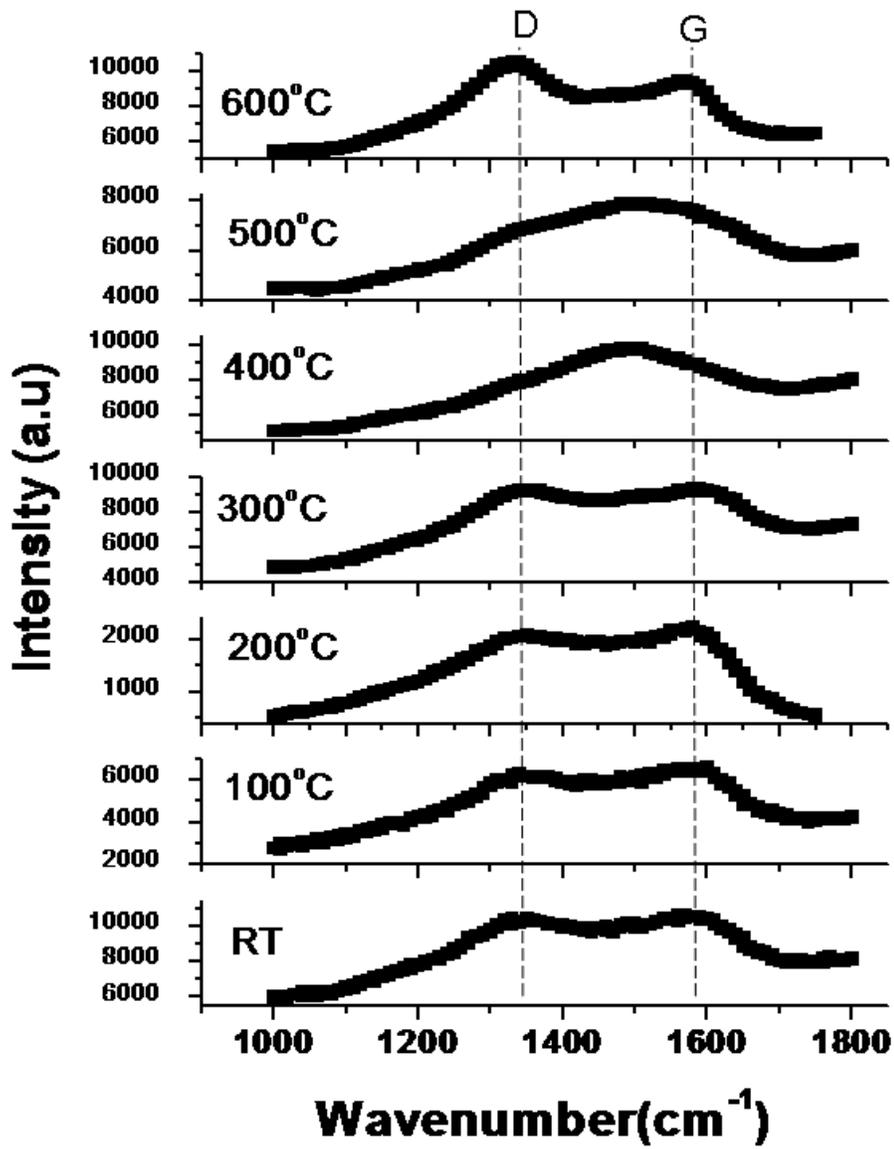

Figure-8

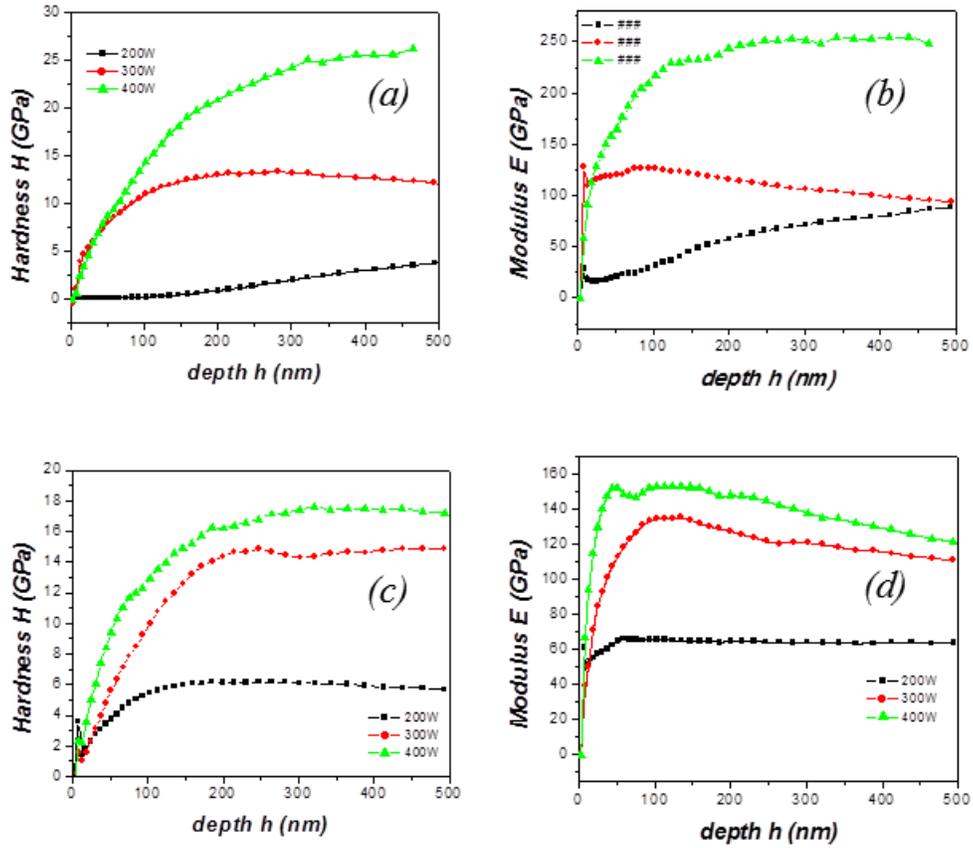

Figure-9

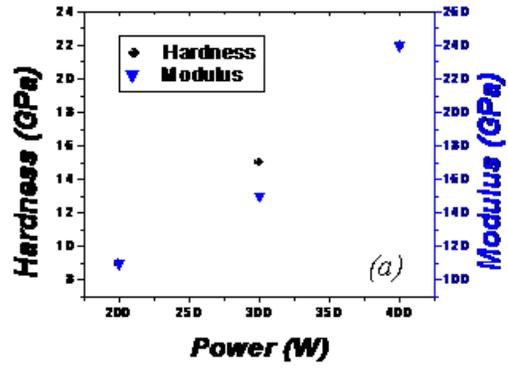

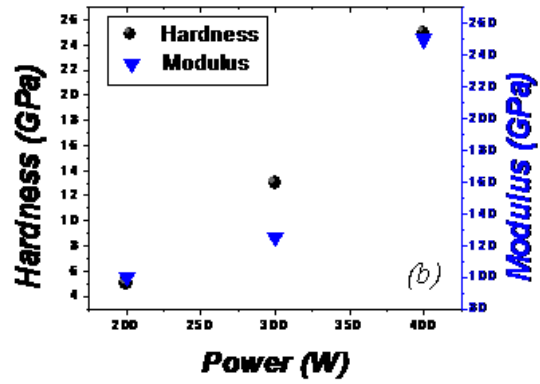

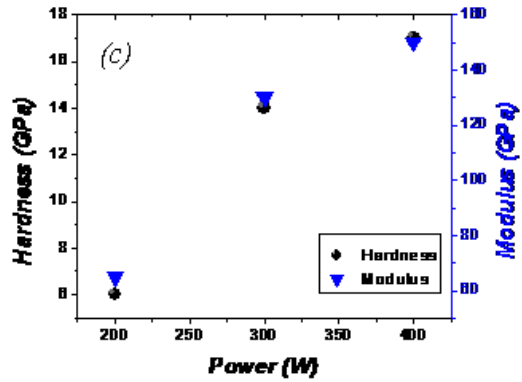

Figure-10

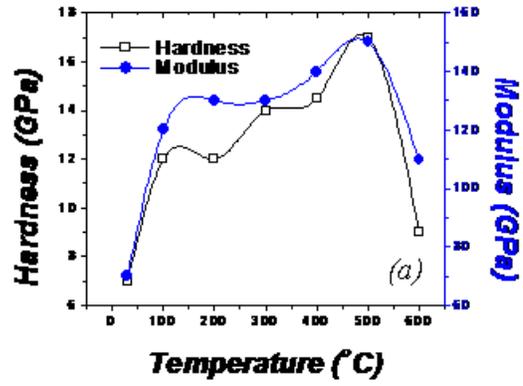

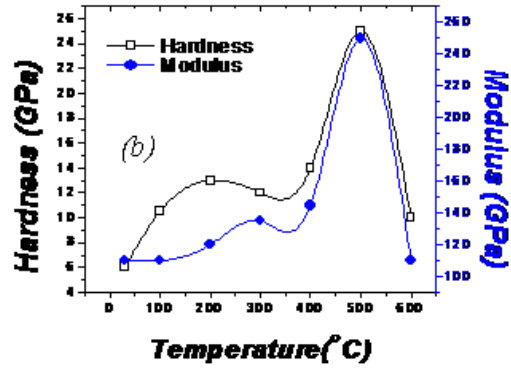

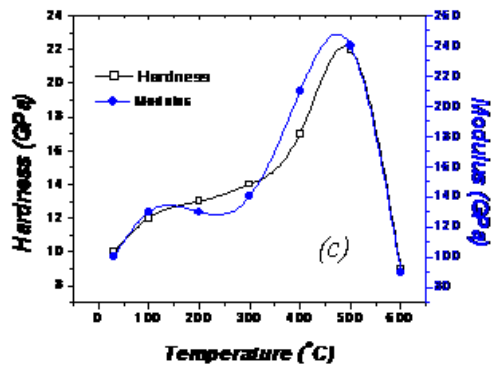

Figure-11

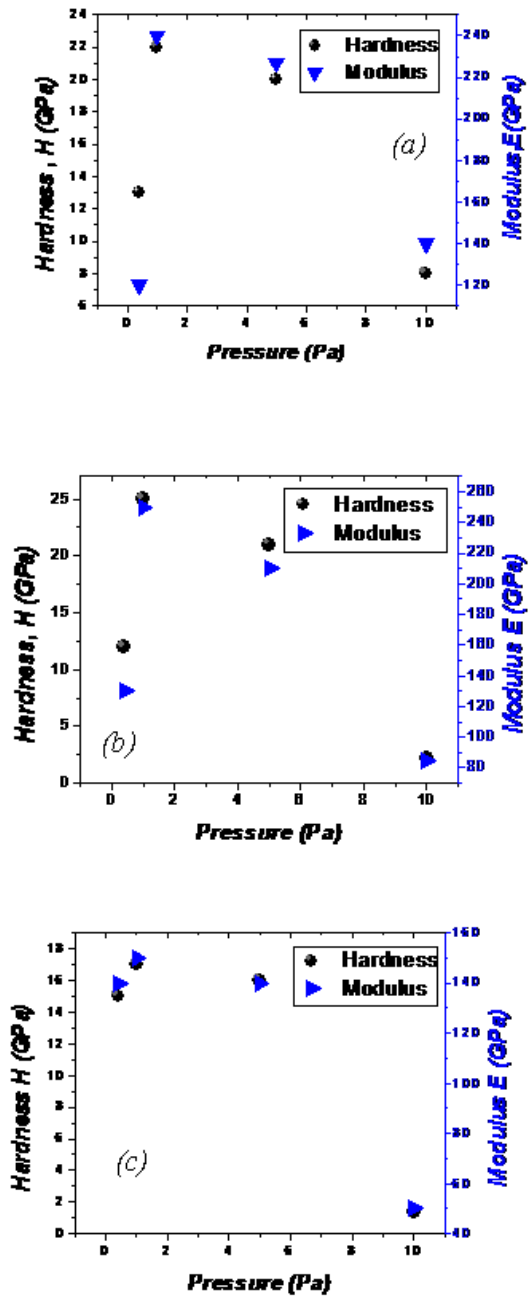

Figure-12

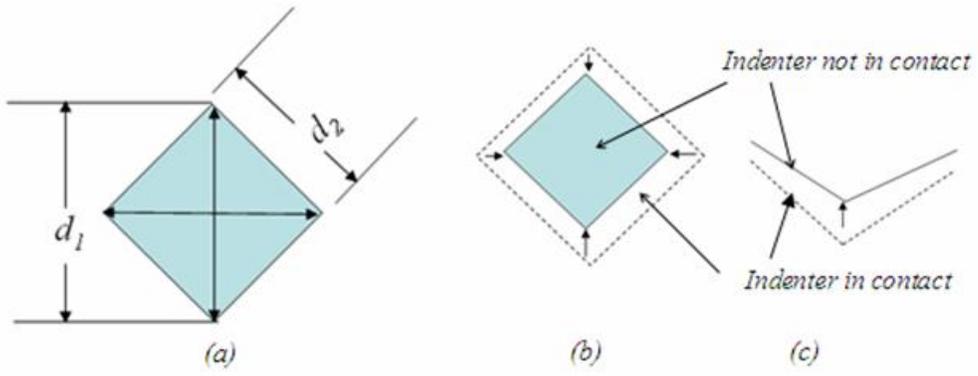

(a)   (b)   (c)

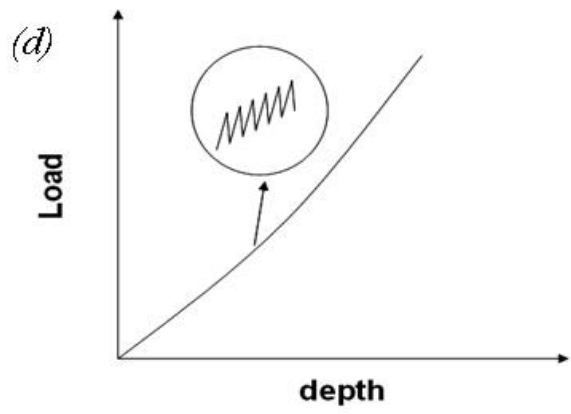

(d)

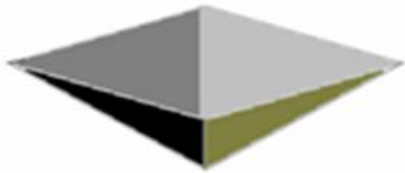

(e)

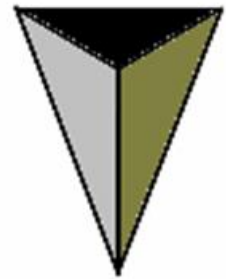

(f)